# 量子相变对三量子比特耦合 XY 自旋链的退纠缠影响

马海福

# 摘要

近年来,随着量子信息这一新兴领域的蓬勃发展,量子纠缠逐渐成为人们的热门话题。量子纠缠是量子力学不同于经典物理的,存在于多子系量子系统中的一种最奇妙、最不可思议的现象。而在实际的量子信息处理过程中,系统不可避免地与环境发生相互作用,从而造成系统的量子相干性破坏或量子态的塌缩,这种现象在量子力学中被称为量子退相干。

本文研究了外场的量子相变对中心粒子退纠缠的影响。本文讨论三量子比特与 XY 自旋链耦合的模型,并研究了其退纠缠的动力学过程。首先对体系的哈密顿量进行求解,找到相干性因子的精确表达式。然后通过对相干性因子及系统参数的讨论,证实量子临界现象对中心自旋粒子退纠缠有极大的影响。并通过分析得到了一个比例规则。

**关键词**:量子比特;XY 自旋链;相干性因子;退纠缠;量子相变

# Abstract

Recently, along with the development of quantum information, quantum entanglemant became a hot topic of people. Quantum entanglemant is one of the most amazing phenomenon in quantum mechanics that is totally different from classical physics. However, system would interact with environment in the practical quantum information process. The entanglement would be broken.

In this paper, we study the disentanglement evolution of three spin qubits in an XY spin-chain environment. The dynamical process of the disentanglement is investigated. We found the exact expression of the coherence factor. We discuss the coherence factor and the parameters, and then we illustrate that the disentanglement of central spins is best enhanced by the quantum critical behavior of the environmental spin chain. Furthermore, a scaling rule is obtained.

# 目录

# 第1章 绪论

## 1.1 研究背景

近年来，随着量子信息这一新兴领域的蓬勃发展，量子纠缠逐渐成为人们的热门话题。量子纠缠是量子力学不同于经典物理的，存在于多子系量子系统中的一种最奇妙、最不可思议的现象，即对一个子系统的测量结果无法独立于其他子系的测量参数。量子纠缠已在量子通讯中得到了广泛地应用：利用纠缠人们创造出了量子纠错编码[1]、稠密编码[2]、隐形传态[3]等经典信息理论不可思议的奇迹；构造出了Shor分解大数质因子[4]、Grover随机数据库搜索[5]等量子算法；利用理想的量子计算机，可以进行大规模的并行计算，实现经典计算机不可比拟的信息处理功能。

量子纠缠现象最早出现在爱因斯坦质疑量子力学体系的完备性时提出的EPR佯谬[6]中，它试图论证在承认局域性和实在性的前提下量子力学描述的不完备性。而量子纠缠的概念和术语是由薛定谔于1935年首次引入量子力学，并称之为"量子力学的精髓"[7]。从此，量子纠缠问题一直是物理学中一个引人注目的研究领域。一方面，量子纠缠体现了量子态的非定域性，有助于人们对量子力学本身基础问题的研究；另一方面，它在量子信息处理，例如量子隐形传态、量子编码及量子纠错、量子密钥分配和量子计算中具有重要的应用价值。可以说，没有对量子纠缠现象的深入研究，就没有今天的量子信息学。早期对于纠缠态的研究都是限于哲学思辨的层次上，直到1964年Bell不等式[8]提出后，才使得量子理论与局域隐变量理论的预言的差别可以通过实验来验证，结果是实验支持了量子力学几率的预言，人们开始把纠缠这一非经典特性应用到信息科学和计算科学中去。

## 1.2 国内外研究现状及发展趋势

在实际的量子信息处理过程中，系统不可避免地与环境发生相互作用，从而造成系统的量子相干性破坏或量子态的塌缩，这种现象在量子力学中被称为量子退相干。由于量子退相干，量子纠缠会发生变化，故研究量子退相干对量子纠缠的影响是一个很有意义的事情。研究系统受外界作用时纠缠的演化规律，我们称之为纠缠动力学。由Zeh[9]在1970年提出、并经Zurek[10]等发展起来的量子退相干理论，比较成功地解释了各种量子退相干现象，他们提出量子退相干是由于系统

与环境之间不可避免的作用造成的。系统的密度矩阵的演化可以用来描述退相干，密度矩阵对角线上的元素代表了经典的概率态，其他非对角元则代表了这些态之间的相干关联。当退相干产生时，体系的密度矩阵迅速对角化，从而使得量子叠加性质消失。

量子退相干的原因主要有以下几点：

1）量子测量造成的退相干。

2）与环境耦合造成的退相干。

3）量子信息的衰减造成的退相干。

近些年来，研究消相干对量子计算的限制以及各种量子系统的消相干过程已经成为量子信息领域中非常重要的内容。其中"两个量子比特与不同环境相互作用时的纠缠演化"[11-16]成为一大热点问题。

最近，人们开始研究外场自旋链中的量子相变（quantum phase transition）对粒子退相干及退纠缠（解纠缠）的影响。本文讨论三量子比特与XY自旋链耦合的情况。我们将首先对体系的哈密顿量进行求解，找到我们关心的函数的精确表达式。然后对此函数展开讨论，研究量子临界现象对中心自旋粒子退纠缠的影响。并通过分析得到了一些较有价值的结论。

# 第 2 章 基础知识

## 2.1 量子比特

现有的经典信息以比特(bit)作为信息单元。从物理的角度来讲，比特是一个两态系统，它可以被制备为两个可识别状态中的一个，如0或1。从物理实现的角度来说，就是要设计一个拥有两个可以区分状态的物理系统，这两种状态之间有很高的能量势垒，以致于它们之间不能进行自发的相互转换。在经典计算机中，电容器平板之间的电压可以用来表示信息比特，有电荷代表1，无电荷代表0。与经典比特相对应，在量子信息学中，量子信息的基本单位是量子比特(qubit)。一般来说，它具有两个态：$|1\rangle$和$|0\rangle$，并且以这两个独立态为基矢，张起一个二维复矢量空间。量子态区别于经典态的最重要的性质在于它的相干叠加性，即它可以表示为上面两种状态的叠加：

$$|\psi\rangle = a|1\rangle + b|0\rangle$$

如果我们对这个量子比特进行测量，那么有$|a|^2$的几率得到1，有$|b|^2$的几率得到0。这里a和b满足$|a|^2 + |b|^2 = 1$。当量子比特处于混合态时，它的状态就不能写成波函数形式而只能用密度矩阵来表示，例如：

$$\rho = \lambda|1\rangle\langle 1| + (\lambda - 1)|0\rangle\langle 0|$$

表示量子比特处于基态和激发态的混合态，其中$0 \leq \lambda \leq 1$。当$\lambda = 0$时，量子比特处于基态；当$\lambda = 1$时，量子比特处于激发态；当$\lambda = \frac{1}{2}$时，量子比特处于最大混合态。

对于N个量子比特，一组合适的基态是由单个量子态$|0\rangle$，$|1\rangle$的$2^N$个直积态给出：

$$|n\rangle = |i_N\rangle \otimes |i_{N-1}\rangle \otimes \ldots \otimes |i_1\rangle,$$

这里$i_k \in 0,1$，并且$n = \sum_{k=0}^{N} i_k 2^{k-1}$。在这个张量积中，最右边的态$|i_1\rangle$属于第一个量子比特。N个量子比特的量子态能表示成这些基态的叠加。例如两个量子比特的情形：

$$|\psi\rangle = a_0|00\rangle + a_1|01\rangle + a_2|10\rangle + a_3|11\rangle,$$

这里系数 $a_k$ 满足归一化条件 $\sum_{k=0}^{N-1}|a_k|^2 = 1$。作用在一个含有N个量子比特的系统的量子操作，用 $N \times N$ 的幺正矩阵表示。

许多两态的量子系统都可以看作量子比特，常见的有:具有正交偏振态的光子、具有自旋的电子或原子核、两能级原子、超导约瑟夫森器件等。量子信息处理的过程就是操控这些量子系统的量子态的演化过程。量子比特的最显著特性是相干叠加性，即量子比特可以处在两个本征态的叠加态上。在对量子比特的操作过程中，两态的叠加振幅可以互相干涉。

## 2.2 量子纠缠

量子纠缠可以说是量子信息的最核心的部分，几乎所有的量子信息的研究都与其有关。那么什么是量子纠缠呢？所谓量子纠缠态是描述多粒子体系或者多自由度体系的一种量子态,这种量子态在任何表象中都不能写成各单粒子或自由度的直积态[17]。按照量子力学的测量理论，对处于 $|\psi\rangle$ 态复合系统子系之一的测量将会使态 $|\psi\rangle$ 坍缩到其中之一项上，从而对子系之一的测量结果瞬间决定了另一个子系的态，这就是所谓的量子纠缠现象[18]，可对其进行如下定义：

当两个子系构成的复合系统处于纯态 $|\psi\rangle$，若 $|\psi\rangle$ 的对偶基展开中含有两项或两项以上（即描述子系的密度算子[19]有两个以上的非零本征值），则称 $|\psi\rangle$ 是一个纠缠态，如果展开式项数等于1，即：

$$|\psi\rangle = |\psi_1\rangle|\psi_2\rangle$$

就称 $|\psi\rangle$ 是非纠缠态（或可分离的）。非纠缠态是两个子系的纯态的直积态。所以反过来也可以定义纠缠态为：复合系统的一个纯态，如果不能写成两个子系统纯态的直积态，这个态就是一个纠缠态。

1951年，Bohm在他的著作[20]中，以自旋为1/2的二粒子为例，对纠缠态给出了更简洁的物理陈述，设1、2两个自旋为1/2的粒子组成的相关体系处于自旋单态，即总自旋为零，这时粒子称为EPR对，并且它们朝相反的方向运动，这个关联体系的状态可以由反对称自旋波函数来描述，即：

$$|\psi\rangle_{1,2} = \frac{1}{\sqrt{2}}\left[\chi_+^1 \chi_-^2 - \chi_-^1 \chi_+^2\right]$$

更一般的情况，粒子处于自旋向上和自旋向下两种状态，1、2两个这样的粒子组成的量子系统可以组成四个纠缠态：

$|0\rangle$ 表示自旋向上的状态，$|1\rangle$ 表示自旋向下的状态

$$|\phi^+\rangle = \frac{1}{\sqrt{2}}(|00\rangle + |11\rangle)$$

$$|\phi^-\rangle = \frac{1}{\sqrt{2}}(|00\rangle - |11\rangle)$$

$$|\psi^+\rangle = \frac{1}{\sqrt{2}}(|01\rangle + |10\rangle)$$

$$|\psi^-\rangle = \frac{1}{\sqrt{2}}(|01\rangle - |10\rangle)$$

通常将这四个态称为Bell基，其中 $|\psi^-\rangle$ 为单态，其余三个态为三重态，这四个态构成四维的希尔伯特空间的一组正交完备归一基。因此，此空间的任何态矢都可以按这4个基展开。

## 2.3 纠缠度量

量子纠缠在量子信息学中占有极其重要的地位，是量子信息和量子计算的资源。那么，如何量化纠缠就被提升到一个很重要的地位。为了表示对纠缠程度的量度，人们引入纠缠度的概念。由于考察角度不同，所引入的纠缠度的定义有好几种，分别有不同的用途，彼此也不完全互相吻合。但是，作为量子纠缠的定量描述，不论定义如何，都满足以下共同准则[21]：

（1）可分离态的纠缠度为零。

（2）对任一子系统进行任何局域幺正变换(LU)不应改变纠缠度，即LU等价的态具有相同的纠缠度。

（3）局域操作并经典通讯(LOCC)下纠缠不增加。

（4）对于直积态，纠缠度应该可加。

这里我们主要介绍两体纠缠的度量方法，它们分别是部分熵纠缠度、形成纠缠度、可提取纠缠度、相对熵纠缠度以及部分转置正定性判据。

1、部分熵纠缠度

对于两体纯态，纠缠度是一个标量，通常用部分熵纠缠度 $E_p$ 来描述一个粒子体系A和B之间的部分熵纠缠度，一个两体纯态 $|\psi\rangle_{AB}$ 的部分熵纠缠度定义为：

$$E_p(|\psi\rangle) = S(\rho_A) = S(\rho_B)$$

其中，$\rho_A = Tr_B \rho_{AB}$ 和 $\rho_B = Tr_A \rho_{AB}$ 是复合系统 $\rho_{AB} = |\psi\rangle_{AB\ AB}\langle\psi|$ 的约化密度矩阵，而 $S(\rho_A)$ 是是von Neumann熵，其定义为：

$$S(\rho_A) = -Tr(\rho_A \log_2 \rho_A)$$

$E_\rho = 0$ 时，说明 $|\psi\rangle_{AB}$ 为非纠缠态；如果 $E_\rho = 1$，则 $|\psi\rangle_{AB}$ 为最大纠缠态，部分熵纠缠度有人也叫做von Neumann 熵。

部分熵纠缠度表征了系统局域的混乱程度，量子的纠缠越厉害，从局部看的不确定越大。与两体纯态相比，两体混态、三体或者多体态的纠缠度的研究就不是那么清楚了，因为多体态的纠缠度包含了更深、更复杂、内容更丰富多彩的特点和性质。

2、形成纠缠度

对于两体量子态 $\rho_{AB}$，形成纠缠度的定义为：

$$E_f(\rho_{AB}) = \min_{\{p_i,|\psi_i\rangle\}} \sum_i p_i E_p |\psi_i\rangle_{AB}$$

$\{p_i, |\psi_i\rangle_{AB}\}$ 是 $\rho_{AB}$ 的的任一分解，即 $\rho_{AB} = \sum_i p_i |\psi_i\rangle\langle\psi_i|$，$|\psi_i\rangle_{AB}$ 可以不是正交的，但是要求 $|\psi_i\rangle_{AB}$ 是两体的归一化纯态，$E_p |\psi_i\rangle_{AB}$ 是部分熵纠缠度。一般两体量子态形成纠缠度的计算并不简单，但是对于两能级体系，两个子体系态空间都是2维的，可以将形成纠缠度直接算出来。引入密度翻转矩阵 $R_{AB}$：

$$R_{AB} \equiv (\sigma_y^A \otimes \sigma_y^B)\rho_{AB}^*(\sigma_y^A \otimes \sigma_y^A)$$

其中，$\sigma_y^A$、$\sigma_y^B$ 是泡利矩阵的分量。算符 $R_{AB}$ 不一定厄米，但半正定。设其本征值为 $\lambda_i^2$ $(i=1,2,3,4)$，且按递减顺序排列，定义Concurrence[22-23]为

$$C(\rho_{AB}) \equiv \max\{0, \lambda_1 - \lambda_2 - \lambda_3 - \lambda_4\}$$

$\rho_{AB}$ 的形成纠缠度为

$$E_F(\rho_{AB}) = H\{\frac{1+\sqrt{1-C^2(\rho_{AB})}}{2}\}$$

其中，

$$H(\rho) = -\rho \log_2 \rho - (1-\rho)\log_2(1-\rho)$$

容易知道 $E_F$ 是Concurrence的单调函数，即 $E_F(\rho_1) = E_F(\rho_2)$ 当且仅当 $C(\rho_1) = C(\rho_2)$。因此，Concurrence可以很好地描述 $2 \times 2$ 系统的纠缠。

3、可提纯纠缠度

Bennett证明通过对n对非最大纠缠对的拷贝进行局域操作和经典通信，可以浓缩它们的纠缠到数目较少的最大纠缠对上。若每个非最大纠缠对的相对熵纠缠度为E，当n足够大时，可以得到最大纠缠对的数目渐近于nE，这个过程称为纠缠的纯化（entanglement purification）或者浓缩（entanglement concentration）。可提纯纠缠度 $E_D$ 可定义为，通过LOCC可以从单位个 $\rho$ 提取出来最大纠缠态的数目。

$$E_D = \lim_{N \to \infty} \frac{k(N)}{N}$$

相反的，从起始的许多个最大纠缠对出发，通过局域操作和经典通信，可以制备出数目更多的非最大纠缠对，这个过程称为纠缠稀释（entanglement dilution）。因此从这种意义上说，纠缠浓缩和纠缠稀释是可逆的。

4、相对熵纠缠度

对于两体量子态 $\rho_{AB}$，相对熵纠缠度定义为态 $\rho_{AB}$ 对于全体可分离态的相对熵的最小值：

$$E_r = \min_{\sigma_{AB} \in D} S(\rho_{AB} \| \sigma_{AB})$$

这里D是两体所有可分离态的集合。$S(\rho_{AB} \| \sigma_{AB})$ 是态 $\rho_{AB}$ 对于可分离态 $\sigma_{AB}$ 的相对熵，定义为：

$$S(\rho_{AB} \| \sigma_{AB}) = Tr[\rho_{AB}(\log \rho_{AB} - \log \sigma_{AB})]$$

在相对熵的定义中，需要考虑两体可分离态的所有情况。通常要寻找使得相对熵达到极小值的 $\sigma_{AB}$，计算上非常困难。

5、部分转置正定性判据

Peres、Horodecki等人证明了，对于由两能级量子比特组成的两体混合系统，部分转置为半正定(PPT)是 $2 \times 2$ 和 $2 \times 3$ 系统可分离的充分必要条件[24-25]，但若维度大于 $2 \times 3$，PPT 仅为必要条件。所以对于一个由密度矩阵 $\rho$ 所描述的 $2 \times 2$ 和 $2 \times 3$ 系统，度量其混合态纠缠度的有效且可以计算的方法是NPT[24-25]，它的定义为：

$$E(\rho) = -2 \sum_i \mu_i^-$$

其中 $\mu_i^-$ 是是密度算符 $\rho$ 的部分转置的负的本征值。$E(\rho)$ 的范围是 $0 \leq E(\rho) \leq 1$，$E(\rho)=1$ 表示系统处于最大纠缠态，$E(\rho)=0$ 表示系统处于可分离态。

上面我们介绍的几种纠缠度的定义只适用于两体系统，而对于一般的多体系统纠缠的度量，现在还是一个难题。特殊情况下，保真度是一个较为合适的选择。

保真度是衡量两个量子态接近程度的一种度量，是量子通讯和量子计算领域中的一个重要的物理量。它不仅能给出两个混态量子态之间的接近程度及量子距离的度量，而且还可以用来度量量子系统的纠缠度。

两个量子态 $\rho$ 和 $\sigma$ 的保真度定义为：

$$F(\rho,\sigma) = Tr\sqrt{\rho^{1/2}\sigma\rho^{1/2}}$$

当其中一个量子态为纯态 $|\Phi\rangle$ 时，保真度为

$$F(\rho,|\Phi\rangle) = Tr\sqrt{\langle\Phi|\rho|\Phi\rangle|\Phi\rangle\langle\Phi|}$$

也即保真度等于 $|\Phi\rangle$ 和 $\rho$ 之间重叠部分的平方根，它揭示了 $|\Phi\rangle$ 和 $\rho$ 之间的相似程度，保真度的范围是 $0 \leq F \leq 1$。当 $F=1$ 时，表示两个态完全相同。特殊地，当我们选择其中一个是最大纠缠态时，我们就可以用保真度来度量纠缠。这为多体量子系统提供了一个有效的方法。文献[26-28]证明 $F \geq 0.5$ 是量子比特处于纠缠态的充分条件。

## 2.4 海森堡自旋链

### 2.4.1 海森堡自旋链模型

海森堡模型[29]是建立在一套假定之上而提出的自旋-自旋相互作用模型。自旋-自旋相互作用系统的哈密顿量通常表示为：

$$H = -\sum_{l,l'}{}' J_{ll'} \cdot \hat{S}_l \cdot \hat{S}_{l'}$$

其中，$\hat{S}_l$ 代表第 $l$ 个格点上磁性离子的矢量自旋算符，$\sum_{l,l'}{}'$ 代表求和时不计 $l=l'$ 项，$J_{ll'}$ 是 $l$ 与 $l'$ 二格点离子上电子间的交换积分，

$$J_{ll'} = J(|l-l'|)$$

上面两式表示的自旋-自旋相互作用模型就是海森堡模型。海森堡相互作用在凝聚态物理中被认为是基本的自旋-自旋相互作用[30]。

在应用海森堡模型时，常采用更为简化的哈密顿量。考虑到交换积分的大小与格点间电子云的重叠程度有关，交换作用是短程作用，可以只计算近邻格点间的作用。若再假定 $J_{l,\,l+\delta}$ 为各向同性的常数 $J$，则海森堡模型表示为

$$H = -J\sum_{l,\delta} \hat{S}_l \cdot \hat{S}_{l+\delta}$$

$\delta$ 代表近邻格点间位置矢差。当 $J>0$ 时，上式的基态为铁磁序；当 $J<0$ 时可用于描述反铁磁性；当 $J>0$ 且近邻格点为不同磁离子($S$ 不同)时，则描述亚铁磁性，即铁淦氧磁序。

海森堡相互作用模型体系的性质，随着体系的几何维度、自旋格点位形、相互作用范围、交换作用参数及格点自旋值等的不同而不同。

体系的几何维度：是自旋晶格体系的空间几何维数，它对体系的性质有着显著的影响。维度不同，体系的性质一般不同。低维体系不仅可以用来描述一些相应的高维实际物质的性质，而且还会有许多丰富而独特的特性。

自旋格点的位形：主要有正方晶格、三角形晶格、六角形晶格及其它相关晶格结构。

相互作用的范围：在绝缘物质中，由于相互作用都是短程的，所进行的研究因此也大多只计及最近邻及次近邻晶格间的相互作用。

交换作用参数：当交换积分 $J$ 为正时，自旋趋于反平行排列而呈现反铁磁性或亚铁磁性；当交换积分 $J$ 为负时，自旋趋于平行排列而呈现铁磁性；如果 $J$ 的大小和符号是变化的，还可以得到螺磁性和其它自旋结构。当晶格的旋转对称性及平移对称性被破坏以后，自旋沿着某一量子化方向的概率就会占优势，因此会引入体系的各向异性，此时 $J$ 为一矢量。对于给定的物质，$J$ 的数值是由实验测得的，可以把它作为一个给定的模型参数。由于电子是局域的，模型已假定在同一格点的离子上，电子间的交换作用可以忽略不计，两格点间所有的电子具有相同的交换积分。

格点自旋值：由于组成体系的原子最外层电子数目的不同，会使体系具有不同的格点自旋值，从而会使体系具有不同的性质。若每个格点的离子上只有一个未配对的局域电子，则格点自旋值 $S=1/2$。

## 2.4.2 自旋1/2

海森堡XYZ链是最普遍的海森堡相互作用模型，海森堡XX，XY，XXX，

XXZ链都是它的特例。位于外部磁场中的自旋为$1/2$的N个量子位的海森堡XYZ模型的哈密顿量[31]是：

$$H = \frac{1}{2}\sum_{i=1}^{N}\left[J_x\sigma_i^x\sigma_{i+1}^x + J_y\sigma_i^y\sigma_{i+1}^y + (B_1\sigma_i^z + B_2\sigma_{i+1}^z)\right]$$

式中$\sigma_i^\alpha(\alpha=x,y,z)$是第$i$个量子位的泡利矩阵，$J_\alpha(\alpha=x,y,z)$是海森堡相互作用强度，磁场$B$的方向沿Z轴方向。对于自旋相互作用，当$J_\alpha > 0$时，海森堡链是反铁磁性的；当$J_\alpha < 0$时，海森堡链是铁磁性的。当$J_x \neq J_y \neq J_z$时，叫XYZ链，当$J_x = J_y = J_z$及$J_x = J_y \neq J_z$时，分别叫XXX和XXZ链，其它类推。当$J_x = J_y$及$J_z = 0$时，叫XX链，也叫各向同性XY链；当$J_x \neq J_y$及$J_z = 0$时叫各向异性XY链。

其中，XY链在量子Hall系统[32]以及腔QED[33]系统中可以实现。当$B_1 = B_2$时，外部磁场为均匀磁场；当$B_1 \neq B_2$时，外部磁场为非均匀的。要实现量子计算，控制单独作用在每一个自旋上的磁场是非常必要的。因此，引入非均匀磁场是非常必要的。

### 2.4.2 自旋 1

具有海森堡交换相互作用的自旋模型，对高自旋系统会出现更复杂的相互作用项，例如：对自旋$S=1$双二次相互作用将会出现。具有双二次作用项的$S=1$反铁磁海森堡链的哈密顿量可以表示为[34]

$$H = J\sum_{i,j}S_i \cdot S_j + \alpha J\sum_{i,j}(S_iS_j)^2 \quad (J>0)$$

具有双二次相互作用项的$S=1$反铁磁海森堡链的哈密顿量可改写成

$$H = J\sum_{i=\frac{N}{2}+1}^{\frac{N}{2}}\left[S_i^zS_{i+1}^z(S_i^+S_{i+1}^- + S_i^-S_{i+1}^+)\right]$$

$$+\alpha J\sum_{i=\frac{N}{2}+1}^{\frac{N}{2}}\left[S_i^zS_{i+1}^z + \frac{1}{2}(S_i^+S_{i+1}^- + S_i^-S_{i+1}^+)\right]$$

式中 $J > 0$，$S_i^x$、$S_i^y$ 和 $S_i^z$ 是格点 $i$ 自旋 $(S=1)$ 的三个分量，$S_i^\pm = S_i^x \pm S_i^y$，$\alpha$ 表示双二次相互作用参数。

作为一种简单而又实际的固态系统，海森堡链可用于量子信息处理。通过适当的编码技术，海森堡链可用做量子计算[35]。

## 2.5 密度算符

### 2.5.1 密度算符

描述系统的态有两种类型，一种是纯态；另一种是统计混合态。如果一个系统的态完全确定，我们就说这种态为纯态，它可由一态矢 $|\psi(t)\rangle$ 描述，可由系统任一物理量的本征态矢集 $|u_n\rangle$ 的叠加态来表示：

$$|\psi(t)\rangle = \sum_n C_n(t)|u_n\rangle$$

式中，$C_n(t) = \langle u_n|\psi(t)\rangle$，$|C_n(t)|^2$ 表征系统处于本征态 $|u_n\rangle$ 的概率，且满足归一化条件：

$$\sum_n |C_n(t)|^2 = 1$$

物理量A在时间t的期望值由态矢 $|\psi(t)\rangle$ 确定：

$$\langle A\rangle = \langle\psi(t)|A|\psi(t)\rangle = \sum_{n,p} C_n^* C_p(t) A_{np}$$

其中，物理量A的矩阵元为：

$$A_{np} = \langle u_n|A|u_p\rangle$$

与期望值相关联的函数 $C_n^* C_p(t)$ 可以看作算符 $|\psi(t)\rangle\langle\psi(t)|$ 的矩阵元：

$$\langle u_p|\psi(t)\rangle\langle\psi(t)|u_n\rangle = C_n^*(t)C_p(t)$$

因此如果引入纯态的密度矩阵算符：

$$\rho(t) = |\psi(t)\rangle\langle\psi(t)|$$

那么，在 $\{|u_n\rangle\}$ 基中密度算符 $\rho(t)$ 的矩阵元可表示为：

$$\rho_{pn}(t) = \langle u_p|\rho_s(t)|u_n\rangle = C_n^*(t)C_p(t)$$

于是，根据我们可以由密度算符给出物理量A的期望值：

$$\langle A \rangle = \sum_{n,p} \langle u_p | \rho(t) | u_n \rangle \langle u_n | A | u_p \rangle$$

$$= \sum_p \langle u_p | \rho(t) A | u_p \rangle$$

$$= Tr[\rho(t) A]$$

如果系统的态不完全确定，设系统处于态$|\psi_1\rangle$的概率为$P_1$，处于态$|\psi_2\rangle$的概率为$P_2$，……。那么我们就说，系统是处于态$|\psi_1\rangle$，$|\psi_2\rangle$……，且概率分别为$P_1$，$P_2$……的统计混合态，这时采用密度算符的方法可以更为方便地描述。密度算符定义为：

$$\rho(t) = \sum_k P_k |\psi_k(t)\rangle \langle \psi_k(t)|$$

此式的定义对量子力学的三个绘景都适用，其中$P_k$为系统处在态$|\psi(t)\rangle$的概率，且在此情况下，物理量A的期望值仍为：

$$\langle A \rangle = Tr[\rho(t) A]$$

对于一个由哈密顿量：

$$H = H_0 + V_s$$

描述的系统，如果$H_0$不显含时间，可以在相互作用绘景中定义态函数$|\psi_k^I(t)\rangle$满足幺正变换：

$$|\psi_k^I(t)\rangle = U(t) |\psi_k(0)\rangle$$

则

$$\rho_I(t) = \sum_k P_k U(t) |\psi_k(0)\rangle \langle \psi_k(0)| U^+(t)$$

$$= U(t) \rho(0) U^+(t)$$

可见，若知道$U(t)$，就可以根据$\rho(0)$得知$\rho_I(t)$。

### 2.5.2 约化密度算符

对于两个彼此有耦合的系统A和B，设系统在$t = 0$时刻以前彼此独立，而且分别有完备基矢集$\{|A'\rangle\}$和$\{|B'\rangle\}$，它们的哈密顿量分别为$H_A$和$H_B$，并且它

们之间满足:

$$[H_A, H_B] = 0$$

若 $t = 0$ 以后，两系统相互耦合，系统的总哈密顿量可写为:

$$H = H_A + H_B + V_I = H_0 + V_I$$

在任一 $t > 0$ 时刻，假设总系统处于密度算子 $\rho_{AB}(t)$ 描述的态，这个态可以是纯态，也可以是统计混合态。如果我们只测量系统A的物理量M，则此时M的期望值为：

$$\begin{aligned}\langle M \rangle &= Tr_{AB}[M\rho_{AB}(t)]\\ &= \sum_{A',B'}\langle A',B'|M\rho_{AB}(t)|A',B'\rangle\\ &= \sum_{A'}\langle A'|M[\sum_{B'}\langle B'|\rho_{AB}(t)|B'\rangle]|A'\rangle\\ &= Tr_A[M\rho_A(t)]\end{aligned}$$

式中，$\rho_A(t)$ 为约化密度矩阵算符，定义为:

$$\rho_A(t) = Tr_B[\rho_{AB}(t)]$$

## 2.6 退相干

量子纠缠给量子信息和量子计算带来了光明灿烂的前景。然而，量子纠缠的脆弱性也给量子信息和量子计算的物理实现带来了障碍。由于环境作用或者其他的原因，使得量子位能量耗散或者相对位相改变，从而导致量子纠缠的消失，这种现象就称为量子退相干。量子退相干可使得相干态成为一个不包含任何量子信息的无价值的态。

研究量子退相干有三种方法：

1) 环境导致的退相干

自20世纪70年代以来，由Zeh提出并由Zurek等发展起来的量子退相干理论，比较成功地解释了各种量子退相干现象。他们认为量子退相干是由于系统和环境之间不可避免的相互作用造成的。这个理论还解释了物体如何由微观体系的叠加态过渡到宏观系统的确定态，并解释了为何世界没有在大尺度下显示叠加性，及世界如何"分裂"等。人们可以借助系统的密度矩阵来描述退相干[36]，因为密度矩阵的对角元素代表了经典的概率态，其它非对角元则代表了这些态之间的相干关

联。当退相干产生时，密度矩阵迅速对角化，从而使得量子叠加性质消失。在这一理论框架下，人们已经进行了大量理论[37]和实验[38]的研究工作，并取得了重要进展。这一理论模型得到了大多数人的支持。

2) 随机退相干

1954年，Anderson[39]和Kubo[40]建立了随机退相干模型。这个模型的根本思想是利用随机过程来描述退相干，具体做法就是在量子力学方程中引入随机变量考虑宏观物体不可避免要与周围环境发生相互作用而引起的随机因素[41]。但是，与环境的相互作用原则上可能尽量减弱，单用环境的影响来解释宏观系统量子态的退相干难以令人信服[42-43]。

2) 内禀退相干

这是一种广泛适用的解释退相干的方法。通过修改Schrödinger方程使相干性在系统达到宏观量级时作为系统的物理特性自动被破坏。有几个建议通过修正薛定谔方程来解决退相干问题[44-50]。在这样一种方法中，当量子系统演化时，量子相干自动被破坏，这意味着在宏观尺度下系统行为是经典的。这种内禀退相干(intrinsic decoherence)方法在几个模型下已被研究。特别地，Milburn[51]基于一个假设对标准量子力学作了简单的修正。这个假设是：在足够短的时间内，系统不做连续的幺正演化而是处于一个相同幺正变换的随机序列中。这样的假设导致修正的薛定谔方程在能量本征态基上包含一项量子相干衰减项，没有与正常衰减相联系的通常的能量耗散。

在标准量子力学中，保守量子系统的动力学由通过演化算符 $U(t) = \exp(-iH/\hbar)$ 支配的密度算符 $\rho$ 来描述，其中H是相应的Hamiltonian量。在时间间隔 $(t, t+\tau)$ 内量子系统态的变化由如下幺正变换支配：

$$\rho(t+\tau) = U(\tau)\rho(t)U^+(\tau)$$
$$= \exp\left[-\frac{i}{\hbar}\tau H\right]\rho(t)\exp\left[\frac{i}{\hbar}\tau H\right]$$

对于任意大小的 $\tau$ 上式都有效。

Milburn 对上述变换式给出了三个新的假设：

(1) 对于足够短的时间步长，系统不依照上述幺正变换连续进行演化，而是随机的变化。系统变化的几率是 $\rho(\tau)$，反映了系统状态的量子跳跃。

(2) 如果系统的状态进行某种变化，那么密度算符按下面关系变化

$$\rho(t+\tau) = \exp\left[-\frac{i}{\hbar}\theta(\tau)H\right]\rho(t)\exp\left[\frac{i}{\hbar}\theta(\tau)H\right]$$

其中 $\theta(\tau)$ 是 $\tau$ 的函数。在标准量子力学中，我们有 $\rho(\tau)=1$ 和 $\theta(\tau)=\tau$。在 Milburn 提出的广义模型中我们只要求对足够大的 $\tau$ 有：

$$\rho(\tau) \to 1 \text{ 和 } \theta(\tau) \to \tau$$

(3) 假设

$$\lim_{\tau \to 0} \theta(\tau) = \theta_0$$

上面引入了一个最小时间步长. 这个时间步长的倒数等于幺正步长的平均频率。

Milburn模型中 $\rho(t)$ 的变化率由下述方程给出：

$$\frac{d}{dt}\rho(t) = \gamma\left\{\exp\left[-\frac{i}{\hbar\gamma}H\right]\rho(t)\exp\left[\frac{i}{\hbar\gamma}H\right] - \rho(t)\right\}$$

其等价于假设在一个很短的时间内系统演化的几率是 $\rho(\tau)=\gamma\tau$。上述是对薛定谔方程的推广.。在 $\gamma \to \infty$（即，当基本时间步长趋于零时），方程约化到通常的密度算符的Von Neumann 方程.。展开方程到 $\gamma^{-1}$ 的一阶项，动力学方程有下列形式：

$$\frac{d}{dt}\rho(t) = -\frac{i}{\hbar}[H,\rho] - \frac{1}{2\hbar^2\gamma}[H,[H,\rho]]$$

方程的一阶修正可使得在能量本征态基矢上密度算符的对角化。再者，这项引入了发散变量，它不和Hamiltonian量对易，但是所有运动常数与Hamiltonian量对易，因此保持不受影响。

## 2.7 量子相变

### 2.7.1 临界现象

物质系统的内部，由于其内部有序度和对称性的差别而引起的、相与相之间在结构、功能、性态等方面的差异，可以用序参量来表征。广义的说，相变是物质系统由一种稳定状态(恒定性态)向另一稳定状态(恒定性态)的跃迁过程。即是指当外场和控制参量连续变化达到某个临界值而引起系统内部对称性的破缺和有序度的突变。临界点是相变现象中的一个关节点，相变系统在临界点邻域表现

出了非常奇特的行为。当控制参量和外场趋近于某个临界点时，系统在微观水平上调整着自身，预示着将出现大的涨落。在临界点，反映系统有序度的序参量连续地出现或消失，某些物理量出现了反常涨落和奇异发散。虽然各种各样的相变系统所含的变量具有很大的差异，但它们的临界行为都显示出极大的相似性，不论其包含的物质和变量如何特殊，在临界点邻域它的变化规律和发展趋势是相同的，这就叫做临界现象[52-53]。

### 2.7.2 量子相变理论

所有宏观物体都是由微观粒子组成的，它们具有量子力学特性的，并按照一定的结构组合在一起。对于一些特殊的微观结构，微观粒子之间会有特定的相互作用，而这些相互作用的整体又决定了这个物体的宏观性质。一般环境下，由于热力学涨落等诸多原因的影响，量子世界的很多奇妙性质会被掩盖而不易让人察觉到。随着低温技术的不断提高，这些被热力学涨落所掩盖的微观量子现象更清晰地向我们展现出来。量子相变(quantum phase transition)现象[54]便是其中之一，它是发生在绝对零度条件下不同于经典相变的另一种相变现象，完全是由量子涨落引起的。量子相变现象在很多物理体系里面都可以发生,如熟知的自旋链系统，狄克(Dick)模型[55]，LMG模型[56]等等。研究表明，一些物质在绝对零度时可以处于不同的相，而且不同的相之间不能够平稳的过渡，物质从一个相转变到另一个相必须要经历一个临界过程。这是因为在不同的条件下，不同种类的量子涨落都试图以各自独特的方式来决定和影响系统的基态。在临界状态的时候，这些不同种类的量子涨落的竞争处于平衡状态。此时，即使一个外部参量或耦合系数生很小的变化，也会引起系统微观特性发生根本性变化，即系统正经历着量子相变。量子相变具体表现在多体系统的基态性质出现一个剧烈变化。在参数空间中，相变的临界点一般为基态能量的奇异点。

与经典相变类似，在量子相变中，关联长度仍是最重要的物理量。在临界点处,关联长度是发散的。由于量子系统中存在经典系统中没有的量子关联—纠缠，人们很自然地想到，可以利用多体系统内部粒子之间的量子纠缠性质来研究量子相变现象。对于强关联系统来说，粒子之间的关联不仅仅限于近邻关联，甚至会有全局关联，因而量子纠缠在这里非常复杂。它应该会反映强关联系统的物理性质，尤其是量子相变。

在2002年，Osterloh及Osborne等人[58]先是研究了一维磁性系统中两近邻粒子之间的量子纠缠在相变点附近的标度律行为，其中的两体纠缠采用了并发度

(Concurrence)[59]来度量。他们的研究表明,量子粒子之间的并发度能够很好的反映整个系统发生的量子相变,并且并发度在临界点附近的标度性质可以用来区分不同的量子相变类型,这个研究揭示了量子非定域性质与量子多体系统性质之间的联系,也激发了人们研究量子纠缠现象与量子多体系统性质之间关系的兴趣。2003年Vidal等人又[60]用块熵(area entropy)研究了自旋系统的量子纠缠,发现其临界行为与共形场论中的熵非常类似,从而建立了凝聚态物理量子场论、和量子信息之间的联系。

# 第 3 章 正文

## 3.1 三量子比特与XY自旋链耦合

本文讨论的模型为三量子比特与一维XY海森堡自旋链耦合。其中三个量子比特为三个相互纠缠的自旋为1/2的中心粒子，处于横向磁场中的XY自旋链为外场。XY自旋链包含参数 $\gamma, \lambda$。本文研究外场自旋链的量子相变对中心三量子比特退纠缠的影响。在参数空间中包含两个临界区域：区间 $(\gamma, \lambda) = (0, (0,1))$，此时外场转化为XX自旋链；临界点 $\lambda_c = 1$ [54]。下面求解此模型。

系统的总哈密顿量是（$\hbar$ 设为单位）：

$$H = -\sum_{l=1}^{N}(\frac{1+\gamma}{2}\sigma_l^x\sigma_{l+1}^x + \frac{1-\gamma}{2}\sigma_l^y\sigma_{l+1}^y + \lambda\sigma_l^z)$$
$$-\frac{g}{2}(\sigma_A^Z + \sigma_B^Z + \sigma_C^Z)\sum_{l=1}^{N}\sigma_l^z \qquad (3.1)$$
$$\equiv H_E^{(\lambda)} + H_I$$

其中公式的第一行定义为 $H_E^{(\lambda)}$，表示自旋链外场的哈密顿量；第二行定义为 $H_I$，表示中心量子比特与自旋链的相互作用。参数 $\lambda$ 表征横向磁场的强度，$\gamma$ 量度面内相互作用的各向异性。当 $\gamma=1$ 时，外场变为 Ising 自旋链，当 $\gamma=0$ 时，外场变为 XX 自旋链。

算符 $(\sigma_A^Z + \sigma_B^Z + \sigma_C^Z)$ 的本征态可以表示为：

$$|1\rangle = |+++\rangle$$
$$|2\rangle = |---\rangle$$
$$|3\rangle = |++-\rangle$$
$$|4\rangle = |+-+\rangle$$

$$|5\rangle = |-++\rangle$$
$$|6\rangle = |+--\rangle$$
$$|7\rangle = |-+-\rangle \quad (3.2)$$
$$|8\rangle = |--+\rangle$$

公式（3.1）中的哈密顿量 H 可以改写为：

$$H = -\sum_l^N \left\{ \frac{1+\gamma}{2}\sigma_l^x\sigma_{l+1}^x + \frac{1-\gamma}{2}\sigma_l^y\sigma_{l+1}^y \right.$$
$$\left. + \left[\lambda + \frac{g}{2}(\sigma_A^Z + \sigma_B^Z + \sigma_C^Z)\right]\sigma_l^z \right\} \quad (3.3)$$

因此 H 又可以改写为：

$$H = \sum_{j=1}^8 |j\rangle\langle j| \otimes H_E^{\lambda_j} \quad (3.4)$$

其中参数 $\lambda_j$ 为：

$$\lambda_1 = \lambda + \frac{3}{2}g, \quad \lambda_2 = \lambda - \frac{3}{2}g$$
$$\lambda_3 = \lambda_4 = \lambda_5 = \lambda + \frac{1}{2}g \quad (3.5)$$
$$\lambda_6 = \lambda_7 = \lambda_8 = \lambda - \frac{1}{2}g$$

$H_E^{\lambda_j}$ 由把 $H_E^{\lambda}$ 中的 $\lambda$ 替换为 $\lambda_j$ 得到。

考虑系统初态 $|\Psi(0)\rangle = |\Psi_S(0)\rangle \otimes |\Psi_E(0)\rangle$，其中 $|\Psi_S(0)\rangle$ 是中心自旋粒子的初态，$|\Psi_E(0)\rangle$ 是外场自旋链的初态。时间演化算子可以写为 U(t)=exp(-iHt)，这时有 $|\Psi(t)\rangle = U(t)|\Psi(0)\rangle$。为求解 U(t)，我们对哈密顿量 H 做 Jordan-Wigner 变换：

$$\sigma_l^x = \prod_{m<l}(1-2a_m^\dagger a_m)(a_l + a_l^\dagger),$$

$$\sigma_l^y = -i\prod_{m<l}(1-2a_m^\dagger a_m)(a_l - a_l^\dagger) \quad (3.6)$$

$$\sigma_l^z = 1 - 2a_l^\dagger a_l$$

经过变换之后，我们得到：

$$H_E^{\lambda_j} = -Nl\sum_l^N [(a_{l+1}^\dagger a_l + a_l^\dagger a_{l+1}) + \gamma(a_{l+1}a_l + a_l^\dagger a_{l+1}^\dagger) \\ + \lambda_j(1-2a_l^\dagger a_l)] \quad (3.7)$$

接下来我们引进傅里叶变换：$d_k = \frac{1}{\sqrt{N}}\sum_l a_l e^{-i2\pi lk/N}$，其中 $k = -M,...,M$，$M = (N-1)/2$。然后再进行 Bogoliubov 变换，哈密顿量可以被对角化，结果为：

$$H_E^{\lambda_j} = \sum_k \Omega_k^{\lambda_j}\left(b_{k,\lambda_j}^\dagger b_{k,\lambda_j} - \frac{1}{2}\right) \quad (3.8)$$

其中能谱 $\Omega_k^{\lambda_j}$ 由下式给出：

$$\Omega_k^{\lambda_j} = 2\sqrt{(\varepsilon_k^{\lambda_j})^2 + \gamma^2 \sin^2 \frac{2\pi k}{N}} \quad (3.9)$$

其中 $\varepsilon_k^{\lambda_j} = \lambda_j - \cos\frac{2\pi k}{N}$，并且

$$b_{k,\lambda_j} = (\cos\frac{\theta_k^{\lambda_j}}{2})d_k - i(\sin\frac{\theta_k^{\lambda_j}}{2})d_{-k}^\dagger \quad (3.10)$$

其中角 $\theta_k^{\lambda_j}$ 满足 $\cos\theta_k^{\lambda_j} = 2\varepsilon_k^{\lambda_j}/\Omega_k^{\lambda_j}$。

哈密顿量（3.4）的时间演化算子由下式给出：

$$U(t) = \sum_{j=1}^8 |j\rangle\langle j| \otimes U_E^{\lambda_j}(t) \quad (3.11)$$

其中 $U_E^{\lambda_j}(t) = \exp(-iH_E^{\lambda_j}t)$。

假设初态自旋粒子 A、B、C 之间是纠缠的，而它们与自旋链之间是非纠缠的。则系统初态可表为直积态：

$$|\Psi_{tot}(0)\rangle = |\phi\rangle_{ABC} \otimes |\psi_E\rangle \qquad (3.12)$$

其中 $|\phi\rangle_{ABC}$ 为三个中心自旋粒子的纠缠初态，$|\psi_E\rangle$ 为外场自旋链的初态。因而，中心自旋粒子演化的约化密度矩阵为：

$$\begin{aligned}\rho_{ABC}(t) &= Tr_E |\Psi_{tot}(t)\rangle\langle\Psi_{tot}(t)| \\ &= \sum_{j,j'=1}^{8} c_j c_{j'}^* \langle\psi_E|U_E^{\dagger\lambda_{j'}}(t)U_E^{\lambda_j}(t)|\psi_E\rangle |j\rangle\langle j'|\end{aligned} \qquad (3.13)$$

其中 $c_j = \langle j|\phi_{ABC}\rangle$。公式（3.13）说明了，外场自旋链仅通过相干性因子 F(t) 调制 $\rho_{ABC}$ 的非对角项：

$$F(t) = \langle\psi_E|U_E^{\dagger\lambda_{j'}}(t)U_E^{\lambda_j}(t)|\psi_E\rangle \qquad (3.14)$$

$\rho_{ABC}$ 的对角项不受外场影响，因为当 $j = j'$ 时，相干性因子为单位值。本文将仅讨论 $j=1$，$j'=2$ 的情形。显然地，$F(t)$ 的值接近 1 表示外场与中心自旋的作用很弱。另一方面，$F(t)$ 的值接近 0，意味着外场自旋链对量子比特的相干和纠缠有很强的抑制作用。

设 $|G\rangle_\lambda$ 为 $H_E^\lambda$ 的基态，它可从 $H_E^{\lambda_j}$ 的基态 $|G\rangle_{\lambda_j}$ 得到：

$$|G\rangle_\lambda = \prod_{k=1}^{M}\left[\cos\alpha_k^{\lambda_j} + i(\sin\alpha_k^{\lambda_j})b_{k,\lambda_j}^\dagger b_{-k,\lambda_j}^\dagger\right]|G\rangle_{\lambda_j} \qquad (3.15)$$

其中 $\alpha_k^{\lambda_j} = (\theta_k^{\lambda_j} - \theta_k^\lambda)/2$。

$F(t)$ 可以被改写为：

$$|F(t)| = \left|{}_\lambda\langle G|U_E^{\dagger\lambda_{j'}}(t)U_E^{\lambda_j}(t)|G\rangle_\lambda\right| \qquad (3.16)$$

下面讨论 $j=1, j'=2$ 的情况，

$$|F(t)| = \left|{}_\lambda\langle G|U_E^{\dagger\lambda_2}(t)U_E^{\lambda_1}(t)|G\rangle_\lambda\right|$$

$$= \Bigg|{}_{\lambda_2}\langle G|\left\{\prod_k\left[\cos\alpha_k^{\lambda_2} - i(\sin\alpha_k^{\lambda_2})b_{-k,\lambda_2}b_{k,\lambda_2}\right]\right\}$$

$$\times e^{iH_E^{\lambda_2}t}e^{-iH_E^{\lambda_1}t}\left\{\prod_k\left[\cos\alpha_k^{\lambda_1} + i(\sin\alpha_k^{\lambda_1})b_{k,\lambda_1}^\dagger b_{-k,\lambda_1}^\dagger\right]\right\}|G\rangle_{\lambda_1}\Bigg|$$

（3.17）

运用等式 $e^{-iH_E^\lambda t}b_{k,\lambda}^\dagger e^{iH_E^\lambda t} = b_{k,\lambda}^\dagger e^{-i\Omega_k^\lambda t}$，公式（17）可以化简为：

$$|F(t)| = \prod_{k>0}\Bigg\{\left[\cos^2\left(\frac{\theta_{\lambda_1}-\theta_{\lambda_2}}{2}\right)\cos\left[(\Omega_{k,\lambda_1}-\Omega_{k,\lambda_2})t\right]\right.$$

$$\left.+\sin^2\left(\frac{\theta_{\lambda_1}-\theta_{\lambda_2}}{2}\right)\cos\left[(\Omega_{k,\lambda_1}+\Omega_{k,\lambda_2})t\right]\right]^2$$

$$+\left[\cos\left(\frac{\theta_{\lambda_1}+\theta_{\lambda_2}}{2}\right)\cos\left(\frac{\theta_{\lambda_1}-\theta_{\lambda_2}}{2}\right)\sin\left[(\Omega_{k,\lambda_1}-\Omega_{k,\lambda_2})t\right]\right.$$

$$\left.-\sin\left(\frac{\theta_{\lambda_1}+\theta_{\lambda_2}}{2}\right)\sin\left(\frac{\theta_{\lambda_1}-\theta_{\lambda_2}}{2}\right)\sin\left[(\Omega_{k,\lambda_1}+\Omega_{k,\lambda_2})t\right]\right]^2\Bigg\}^{1/2}$$

（3.18）

此式即为相干性因子的严格解。根据此式，我们下面将展开一些讨论。

## 3.2 自旋链的量子相变对中心粒子退纠缠的影响

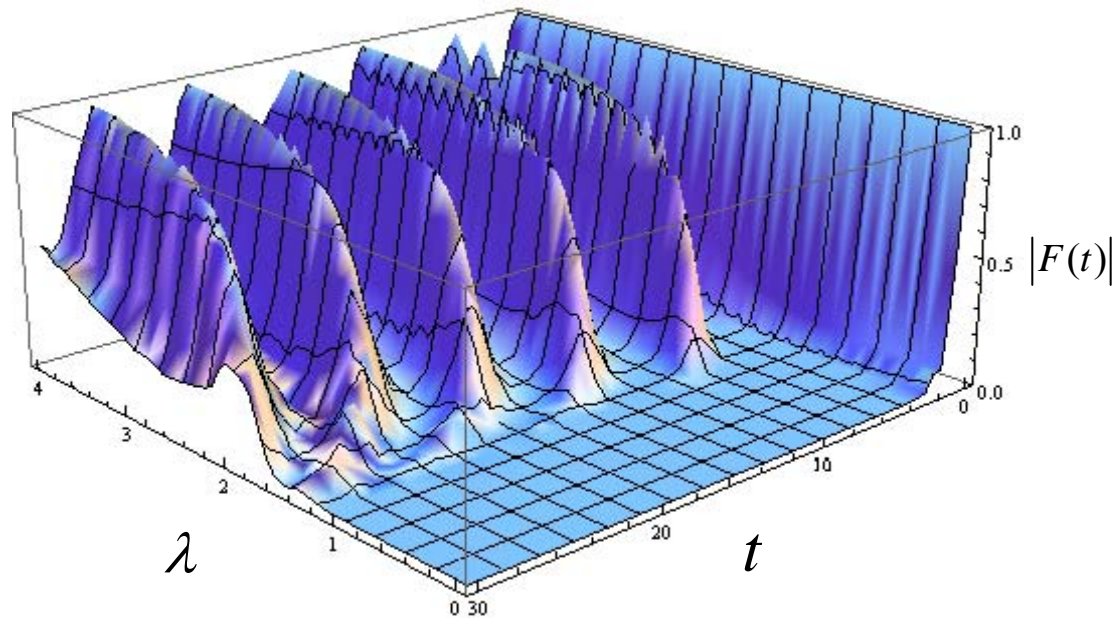

图 3.1

图 3.1 中，|F(t)|是磁场强度 $\lambda$ 和时间 t 的函数。其中，N=101，g=0.1，$\gamma$ =1.0，即模型变为 Ising 模型，且为弱耦合状态。从图中可以看到，在远离临界点 $\lambda_c$ =1 的地方，|F(t)|表现出震荡行为。当 $\lambda$ 趋近 $\lambda_c$ 时，震荡迅速衰减。根本性的变化发生在量子相变的临界点：$\lambda = \lambda_c = 1$。图中显示，在临界点，|F(t)|很快地演化为 0。这说明外场自旋链的量子相变极大地促进了中心自旋粒子的退纠缠。

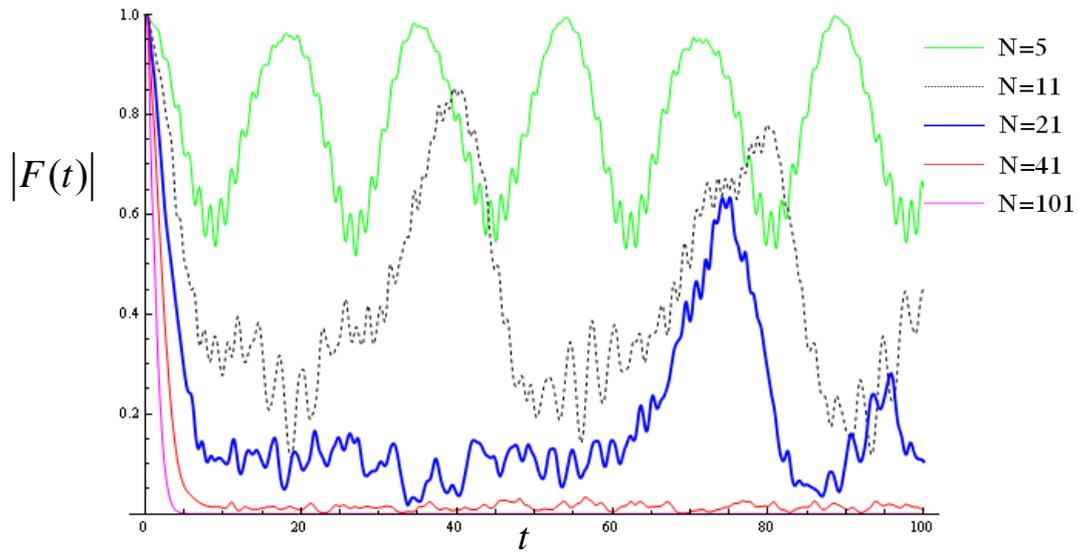

图 3.2

图 3.2 显示了自旋链大小的不同在量子相变点对相干性因子的影响，其中 g=0.05，$\lambda=1$，$\gamma=1$。当 N 趋近热力学极限时，量子相变在 Ising 自旋链里的作用变得非常明显，它在很短的时间内就完全解开了三个中心量子比特的纠缠。

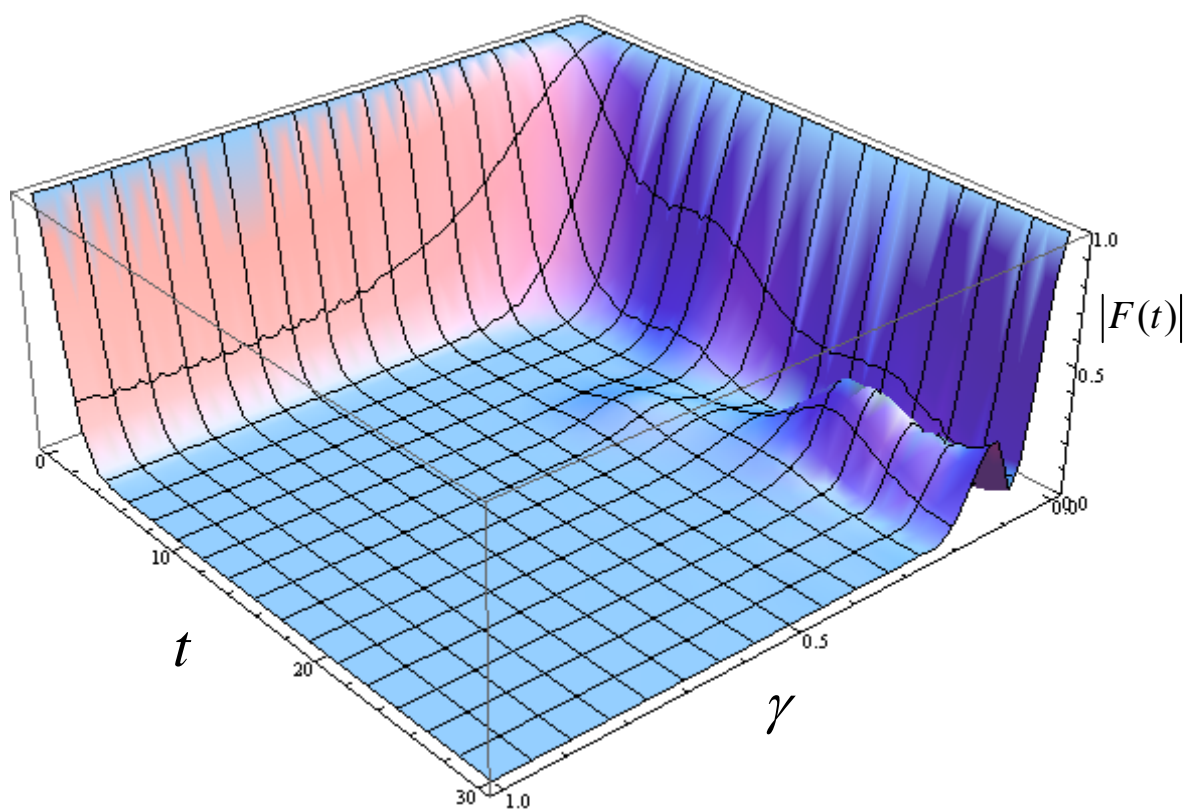

图 3.3

图 3.3 中，相干性因子|F(t)|是自旋各向异性参数 $\gamma$ 和时间 t 的函数。模型为三个量子比特与 XY 自旋链耦合，其中 $\lambda = 1$，N=101，g=0.05 。

我们注意到，当 $\gamma = 0$ 时，|F(t)|不随时间演化，始终为 1。此时，外场自旋链的量子相变对中心自旋粒子的退纠缠没有作用。因此，量子相变对退纠缠的促进，可能由于自旋链模型中各向异性参数特殊的选择而被破坏。

从图中可以看出，当 $\gamma$ 远离 0 的时候，相干性因子震荡地逐渐演化为 0。

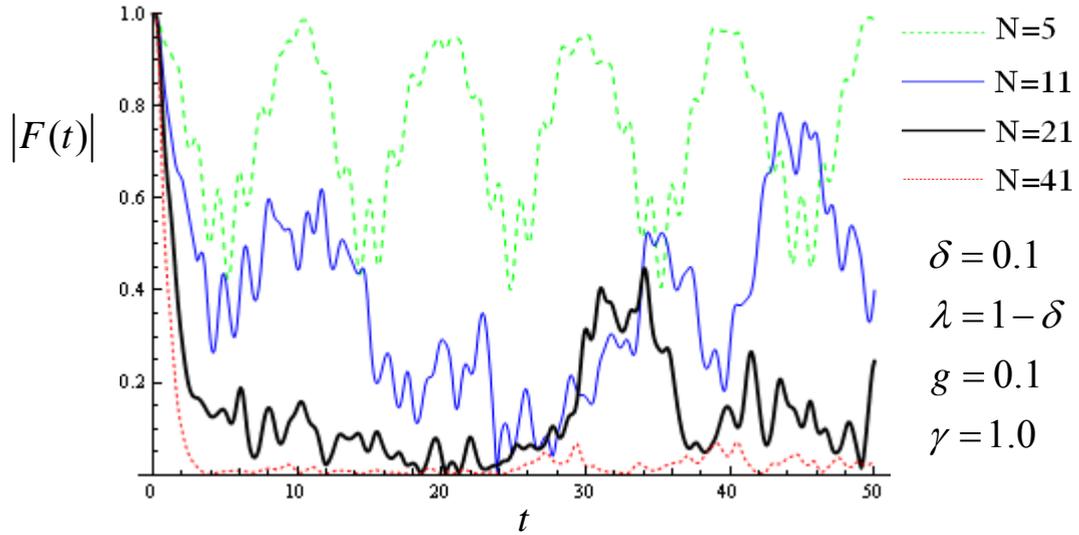

图 3.4

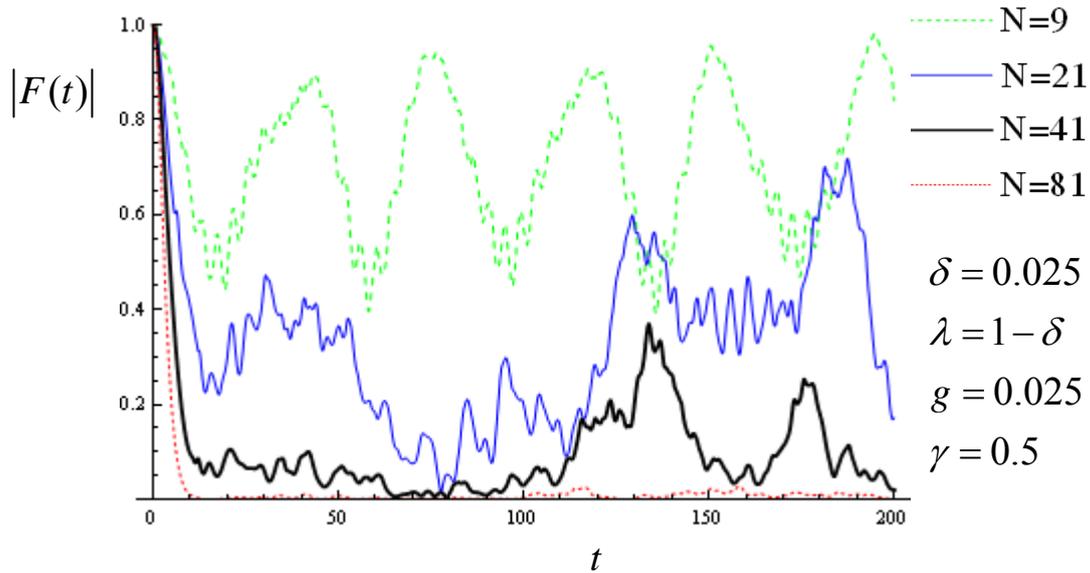

图 3.5

我们发现，在量子相变点附近，相干性因子|F(t)|随时间演化时，其在下列的变换下保持不变：

$$t \to mt, \delta \to \delta/m, g \to g/m, \gamma/N \to \gamma/(mN)$$

其中 $\delta = \lambda_c - \lambda$ 描绘量子相变点附近的特征。为了阐述这种值得注意的比例性质，我们在图 3.4 和图 3.5 中画出了|F(t)|在不同系统参数下随时间演化的过程。其中图 3.5 的参数值按 m=4 的比例从图 3.4 中得到。这两图证明|F(t)|随时间的演化确实遵循比例转换规则。

# 结论

  本文研究了三量子比特与 XY 自旋链耦合时的退纠缠动力学过程，得到了相干性因子|F(t)|的精确表达形式。广泛讨论了|F(t)|与外场自旋链的量子相变的关系。一般的，三量子比特的退纠缠在外场自旋链发生量子相变的时候被最大的增强。分析表明，相干性因子急剧衰减的行为在临界点$\lambda = \lambda_c = 1$的附近。这种衰减行为在$\gamma = 0$（即 XX 自旋链）的特殊情况下被破坏，此时|F(t)|不受外场影响。进一步，我们为相干性因子在量子相变点附近的演化建立了一套比例规则。我们期待这种比例规则可以用来引导将来对多量子比特的退纠缠动力学的实验测量。

# 参考文献

# 附录

用于画图的 Mathematica 程序。

一、图 1 的程序。

```
g = 0.1;
γ = 1;
M = 101;
e = λ - Cos[2 π * k / M];
e1 = λ + 1.5 g - Cos[2 π * k / M];
e2 = λ - 1.5 g - Cos[2 π * k / M];
Ω = 2 √( e^2 + γ^2 * (Sin[2 π * k / M])^2 );
Ω1 = 2 √( e1^2 + γ^2 * (Sin[2 π * k / M])^2 );
Ω2 = 2 √( e2^2 + γ^2 * (Sin[2 π * k / M])^2 );
θ = ArcCos[2 e / Ω];
θ1 = ArcCos[2 e1 / Ω1];
θ2 = ArcCos[2 e2 / Ω2];
a1 = Cos[(θ1 - θ2) / 2];
a2 = Cos[(θ1 + θ2) / 2];
a3 = Sin[(θ1 - θ2) / 2];
a4 = Sin[(θ1 + θ2) / 2];
b1 = Cos[(Ω1 - Ω2) t];
b2 = Cos[(Ω1 + Ω2) t];
b3 = Sin[(Ω1 - Ω2) t];
b4 = Sin[(Ω1 + Ω2) t];
Plot3D[
 Product[((a1^2 * b1 + a3^2 * b2)^2 + (a2 * a1 * b3 - a4 * a3 * b4)^2)^0.5,
  {k, 1, (M - 1) / 2}], {λ, 0, 4}, {t, 0, 30}, PlotRange → {0, 1}]
```

二、图 2 的程序（图 4、图 5 的程序类似）。

```
g = 0.05;
γ = 1;
M = 5;
λ = 1;
e = λ - Cos[2 π * k / M];
e1 = λ + 1.5 g - Cos[2 π * k / M];
e2 = λ - 1.5 g - Cos[2 π * k / M];
Ω = 2 √( e^2 + γ^2 * (Sin[2 π * k / M])^2 );
Ω1 = 2 √( e1^2 + γ^2 * (Sin[2 π * k / M])^2 );
Ω2 = 2 √( e2^2 + γ^2 * (Sin[2 π * k / M])^2 );
θ = ArcCos[2 e / Ω];
θ1 = ArcCos[2 e1 / Ω1];
θ2 = ArcCos[2 e2 / Ω2];
a1 = Cos[(θ1 - θ2) / 2];
a2 = Cos[(θ1 + θ2) / 2];
```

```
a3 = Sin[(θ1 - θ2) / 2];
a4 = Sin[(θ1 + θ2) / 2];
b1 = Cos[(Ω1 - Ω2) t];
b2 = Cos[(Ω1 + Ω2) t];
b3 = Sin[(Ω1 - Ω2) t];
b4 = Sin[(Ω1 + Ω2) t];
N1 =
 Plot[Product[((a1^2*b1 + a3^2*b2)^2 + (a2*a1*b3 - a4*a3*b4)^2)^0.5,
    {k, 1, (M - 1) / 2}], {t, 0, 100}, PlotRange -> {0, 1},
   PlotStyle -> {RGBColor[0, 1, 0]}]
g = 0.05;
γ = 1;
M = 11;
λ = 1;
e = λ - Cos[2 π * k / M];
e1 = λ + 1.5 g - Cos[2 π * k / M];
e2 = λ - 1.5 g - Cos[2 π * k / M];
Ω = 2 √(e^2 + γ^2 * (Sin[2 π * k / M])^2) ;
Ω1 = 2 √(e1^2 + γ^2 * (Sin[2 π * k / M])^2) ;
Ω2 = 2 √(e2^2 + γ^2 * (Sin[2 π * k / M])^2) ;
θ = ArcCos[2 e / Ω];
θ1 = ArcCos[2 e1 / Ω1];
θ2 = ArcCos[2 e2 / Ω2];
a1 = Cos[(θ1 - θ2) / 2];
a2 = Cos[(θ1 + θ2) / 2];
a3 = Sin[(θ1 - θ2) / 2];
a4 = Sin[(θ1 + θ2) / 2];
b1 = Cos[(Ω1 - Ω2) t];
b2 = Cos[(Ω1 + Ω2) t];
b3 = Sin[(Ω1 - Ω2) t];
b4 = Sin[(Ω1 + Ω2) t];
N2 =
 Plot[Product[((a1^2*b1 + a3^2*b2)^2 + (a2*a1*b3 - a4*a3*b4)^2)^0.5,
    {k, 1, (M - 1) / 2}], {t, 0, 100}, PlotRange -> {0, 1},
   PlotStyle -> {RGBColor[0, 0, 0], Dashing[0.005]}]
g = 0.05;
γ = 1;
M = 21;
λ = 1;
e = λ - Cos[2 π * k / M];
e1 = λ + 1.5 g - Cos[2 π * k / M];
e2 = λ - 1.5 g - Cos[2 π * k / M];
Ω = 2 √(e^2 + γ^2 * (Sin[2 π * k / M])^2) ;
Ω1 = 2 √(e1^2 + γ^2 * (Sin[2 π * k / M])^2) ;
Ω2 = 2 √(e2^2 + γ^2 * (Sin[2 π * k / M])^2) ;
```

```
θ = ArcCos[2 e / Ω];
θ1 = ArcCos[2 e1 / Ω1];
θ2 = ArcCos[2 e2 / Ω2];
a1 = Cos[(θ1 - θ2) / 2];
a2 = Cos[(θ1 + θ2) / 2];
a3 = Sin[(θ1 - θ2) / 2];
a4 = Sin[(θ1 + θ2) / 2];
b1 = Cos[(Ω1 - Ω2) t];
b2 = Cos[(Ω1 + Ω2) t];
b3 = Sin[(Ω1 - Ω2) t];
b4 = Sin[(Ω1 + Ω2) t];
N3 =
 Plot[Product[((a1^2 * b1 + a3^2 * b2)^2 + (a2 * a1 * b3 - a4 * a3 * b4)^2)^0.5,
    {k, 1, (M - 1) / 2}], {t, 0, 100}, PlotRange → {0, 1},
   PlotStyle -> {RGBColor[0, 0, 1], AbsoluteThickness[1.8]}]
g = 0.05;
γ = 1;
M = 41;
λ = 1;
e = λ - Cos[2 π * k / M];
e1 = λ + 1.5 g - Cos[2 π * k / M];
e2 = λ - 1.5 g - Cos[2 π * k / M];
Ω = 2 √(e^2 + γ^2 * (Sin[2 π * k / M])^2) ;
Ω1 = 2 √(e1^2 + γ^2 * (Sin[2 π * k / M])^2) ;
Ω2 = 2 √(e2^2 + γ^2 * (Sin[2 π * k / M])^2) ;
θ = ArcCos[2 e / Ω];
θ1 = ArcCos[2 e1 / Ω1];
θ2 = ArcCos[2 e2 / Ω2];
a1 = Cos[(θ1 - θ2) / 2];
a2 = Cos[(θ1 + θ2) / 2];
a3 = Sin[(θ1 - θ2) / 2];
a4 = Sin[(θ1 + θ2) / 2];
b1 = Cos[(Ω1 - Ω2) t];
b2 = Cos[(Ω1 + Ω2) t];
b3 = Sin[(Ω1 - Ω2) t];
b4 = Sin[(Ω1 + Ω2) t];
N4 =
 Plot[Product[((a1^2 * b1 + a3^2 * b2)^2 + (a2 * a1 * b3 - a4 * a3 * b4)^2)^0.5,
    {k, 1, (M - 1) / 2}], {t, 0, 100}, PlotRange → {0, 1},
   PlotStyle -> {RGBColor[1, 0, 0]}]
g = 0.05;
γ = 1;
M = 101;
λ = 1;
```

```
e = λ - Cos[2 π * k / M];
e1 = λ + 1.5 g - Cos[2 π * k / M];
e2 = λ - 1.5 g - Cos[2 π * k / M];
Ω = 2 √(e^2 + γ^2 * (Sin[2 π * k / M])^2) ;
Ω1 = 2 √(e1^2 + γ^2 * (Sin[2 π * k / M])^2) ;
Ω2 = 2 √(e2^2 + γ^2 * (Sin[2 π * k / M])^2) ;
θ = ArcCos[2 e / Ω];
θ1 = ArcCos[2 e1 / Ω1];
θ2 = ArcCos[2 e2 / Ω2];
a1 = Cos[(θ1 - θ2) / 2];
a2 = Cos[(θ1 + θ2) / 2];
a3 = Sin[(θ1 - θ2) / 2];
a4 = Sin[(θ1 + θ2) / 2];
b1 = Cos[(Ω1 - Ω2) t];
b2 = Cos[(Ω1 + Ω2) t];
b3 = Sin[(Ω1 - Ω2) t];
b4 = Sin[(Ω1 + Ω2) t];
N5 =
 Plot[Product[((a1^2 * b1 + a3^2 * b2)^2 + (a2 * a1 * b3 - a4 * a3 * b4)^2)^0.5,
   {k, 1, (M - 1) / 2}], {t, 0, 100}, PlotRange → {0, 1},
  PlotStyle -> {RGBColor[1, 0, 1]}]
N10 = Show[N1, N2, N3, N4, N5]
```

三、图3的程序

```
g = 0.05;
λ = 1;
M = 101;
e = λ - Cos[2 π * k / M];
e1 = λ + 1.5 g - Cos[2 π * k / M];
e2 = λ - 1.5 g - Cos[2 π * k / M];
Ω = 2 √(e^2 + γ^2 * (Sin[2 π * k / M])^2) ;
Ω1 = 2 √(e1^2 + γ^2 * (Sin[2 π * k / M])^2) ;
Ω2 = 2 √(e2^2 + γ^2 * (Sin[2 π * k / M])^2) ;
θ = ArcCos[2 e / Ω];
θ1 = ArcCos[2 e1 / Ω1];
θ2 = ArcCos[2 e2 / Ω2];
a1 = Cos[(θ1 - θ2) / 2];
a2 = Cos[(θ1 + θ2) / 2];
a3 = Sin[(θ1 - θ2) / 2];
a4 = Sin[(θ1 + θ2) / 2];
b1 = Cos[(Ω1 - Ω2) t];
b2 = Cos[(Ω1 + Ω2) t];
b3 = Sin[(Ω1 - Ω2) t];
b4 = Sin[(Ω1 + Ω2) t];
Plot3D[Product[((a1^2 * b1 + a3^2 * b2)^2 + (a2 * a1 * b3 - a4 * a3 * b4)^2)^0.5,
   {k, 1, (M - 1) / 2}], {γ, 0, 1}, {t, 0, 30}, PlotRange → {0, 1}]
```